\documentclass{article}
\usepackage{spconf,amsmath,graphicx}
\usepackage{times}
\usepackage{graphicx}
\usepackage{amsmath}
\usepackage{amssymb}
\usepackage{algorithm}
\usepackage{tikz}
\usepackage{subfig}
\usetikzlibrary{spy,calc}
\usepackage{algorithmic}
\usepackage{enumerate}
\usepackage{multirow}
\usepackage{multicol}
\usepackage[font=footnotesize,labelformat=simple]{subcaption}
\usepackage{arydshln}
\usepackage{tikz}
\usetikzlibrary{spy}
\usepackage[hyphens,spaces,obeyspaces]{url}
\usepackage[colorlinks,allcolors=blue]{hyperref}
\newif\ifblackandwhitecycle

\usepackage{color}
\usepackage{amssymb}
\def\0{{\mathbf 0}}
\def\1{{\mathbf 1}}

\def\f{{\mathbf f}}

\def\l{{\mathbf l}}

\def\p{{\mathbf p}}

\def\r{{\mathbf r}}

\def\v{{\mathbf v}}

\def\x{{\mathbf x}}
\def\y{{\mathbf y}}

\def\B{{\mathbf B}}

\def\D{{\mathbf D}}

\def\I{{\mathbf I}}

\def\L{{\mathbf L}}
\def\M{{\mathbf M}}

\def\S{{\mathbf S}}

\def\U{{\mathbf U}}
\def\V{{\mathbf V}}
\def\W{{\mathbf W}}

\def\ie{{\textit{i.e.}}}

\def\cE{{\mathcal E}}

\def\cG{{\mathcal G}}

\def\cL{{\mathcal L}}

\def\cN{{\mathcal N}}
\def\cO{{\mathcal O}}

\def\cV{{\mathcal V}}

\def\bPsi{{\boldsymbol \Psi}}
\def\bSigma{{\boldsymbol \Sigma}}

\newtheorem{theorem}{\textbf{Theorem}}

\newcommand{\comment}[1]{}

\title{Constructing an Interpretable Deep Denoiser by Unrolling Graph Laplacian Regularizer}
\name{{Seyed Alireza Hosseini$^\dag$, Tam Thuc Do$^\dag$, Gene Cheung$^\dag$, Yuichi Tanaka$^\star$}
\thanks{The work of G. Cheung was supported in part by the Natural Sciences and Engineering Research Council of Canada (NSERC) RGPIN-2019-06271, RGPAS-2019-00110.}}
\address{$^\dag$York University, Canada ~~~~~~ $^\star$Osaka University, Japan}
\begin{document}
%
\maketitle
\begin{abstract}
An image denoiser can be used for a wide range of restoration problems via the Plug-and-Play (PnP) architecture. 
In this paper, we propose a general framework to build an interpretable graph-based deep denoiser (GDD) by unrolling a solution to a \textit{maximum a posteriori} (MAP) problem equipped with a graph Laplacian regularizer (GLR) as signal prior. 
Leveraging a recent theorem showing that any (pseudo-)linear denoiser $\bPsi$, under mild conditions, can be mapped to a solution of a MAP denoising problem regularized using GLR, we first initialize a graph Laplacian matrix $\L$ via truncated Taylor Series Expansion (TSE) of $\bPsi^{-1}$. 
Then, we compute the MAP linear system solution by unrolling iterations of the conjugate gradient (CG) algorithm into a sequence of neural layers as a feed-forward network---one that is amenable to parameter tuning. 
The resulting GDD network is ``graph-interpretable", low in parameter count, and easy to initialize thanks to $\L$ derived from a known well-performing denoiser $\bPsi$. 
Experimental results show that GDD achieves competitive image denoising performance compared to competitors, but employing far fewer parameters, and is more robust to covariate shift.
\end{abstract}
\section{Introduction}
\label{sec:intro}
Denoising is the oldest and most well-studied image restoration problem in the literature. 
A well-designed denoiser can also be used for other restoration tasks, such as interpolation \cite{chen21} and deblurring \cite{bai19}, where the denoiser is used to solve a sub-problem during an \textit{alternating directional multiplier method} (ADMM) iteration \cite{boyd11}, under the \textit{Plug-and-Play} (PnP) framework \cite{chan17}.
While many early denoisers are based on mathematical models such as \textit{bilateral filter} (BF) \cite{tomasi98}, sparse coding \cite{elad06}, low-rank matrices \cite{dong13,dong15}, and BM3D \cite{dabov07,maggioni13}, recent advances in \textit{deep learning} (DL) have led to state-of-the-art (SOTA) denoisers such as DnCNN \cite{zhang17}. 
However, DL solutions based on generic network architectures, such as convolutional neural nets (CNN), generative adversarial nets (GAN) and transformers, are difficult to initialize, require big data volume for network parameter training, and behave like ``black boxes" that are not easy to interpret. 

As an alternative approach, \textit{algorithm unrolling} \cite{monga21} implements model-based algorithm iterations as neural layers of a feed-forward network (FFN) for data-driven parameter optimization, resulting in an algorithm-specific neural net that is interpretable and robust, while requiring a smaller training set to tune fewer parameters.
A classic example is the unrolling of \textit{iterative soft-thresholding algorithm} (ISTA) in sparse coding into \textit{Learned} ISTA (LISTA) \cite{gregor10}, with results showing that, with end-to-end parameter tuning, better denoising performance can be achieved using fewer layers (iterations). 

The crux in algorithm unrolling is to decide on \textit{which} model-based algorithm to unroll, so that the resulting unrolled FFN has good performance along with desirable properties.
In this paper, leveraging recent progress in \textit{graph signal processing} (GSP) \cite{ortega18ieee,cheung18}, we build an FFN unrolled from a graph-based algorithm that can be initialized from a known well-performing denoiser $\bPsi$, so that \textbf{the unrolled FFN is guaranteed a quality benchmark before parameter tuning further improves denoising performance.} 

In GSP, a \textit{maximum a posteriori} (MAP) image restoration problem typically employs a signal prior such as the \textit{graph Laplacian regularizer} (GLR) \cite{pang17} $\x^\top \L \x$, where the sought signal $\x$ is assumed smooth with respect to (w.r.t.) a graph $\cG$ specified by graph Laplacian matrix $\L$. 
GLR has been successfully applied to a range of image restoration problems, including image denoising \cite{pang17}, JPEG dequantization \cite{liu17}, contrast enhancement \cite{yeganeh23}, and point cloud denoising \cite{dinesh20}.
Unrolling of a GLR-regularized denoising algorithm was studied for images \cite{zeng19}, while unrolling of a \textit{graph total variation} (GTV)-regularized denoising algorithm was studied for 2D images \cite{vu21} and light field images \cite{yoshida22}. 

Previous unrolling of GLR/GTV \cite{zeng19,vu21,yoshida22} is complicated by several factors.
First, the matrix inverse operation employed to solve a linear system is computation-intensive and not amenable to back-propagation needed to optimize parameters end-to-end.
Second, like previous DL networks \cite{zhang17}, parameter initialization remains a basic challenge; typically random or all-zero initialization is used, which does not guarantee good performance, given the non-convex nature of the parameter tuning problem and the local optimality of the commonly used \textit{stochastic gradient descent} (SGD) procedure to learn these parameters \cite{Goodfellow-et-al-2016}. 

In this paper, leveraging a recent Theorem 1 in \cite{viswarupan2023mixed} stating that \textit{any} (pseudo-)linear denoiser $\bPsi$, under mild conditions (symmetric, positive definite, non-expansive), is a solution filter to the MAP denoising problem regularized by GLR, we construct an interepretable deep denoiser---called \textit{graph-based deep denoiser} (GDD)---via algorithm unrolling, one that can be initialized using a known good denoiser $\bPsi$ such as BF.
Specifically, given $\bPsi$, we first initialize a graph filter by computing the corresponding graph Laplacian $\L$ via Theorem 1 \cite{viswarupan2023mixed} and truncated \textit{Taylor Series Expansion} (TSE).
$\L$ provides an insightful graph interpretation of $\bPsi$: each entry $L_{i,j}$ encodes the expected similarity between pixels $i$ and $j$.

Next, to mitigate matrix inverse operations, we solve linear system $(\I + \mu \L) \x^* = \y$ for denoised output $\x^*$ by unrolling the \textit{conjugate gradient} (CG) algorithm \cite{stiefel52}, resulting in an FFN that is amenable to end-to-end parameter tuning \cite{thuc23}.
Unlike \cite{zeng19,vu21,yoshida22}, our FFN is initialized using trusted denoiser $\bPsi$ with known good performance, and only improves thereafter via SGD.
Further, GDD is 100\% interpretable---each layer corresponds to one CG iteration---unlike black-box-like denoising networks such as \cite{zhang17}.
Experimental results show that our proposed GDD is competitive with SOTA denoising schemes \cite{zhang17}, while requiring much fewer network parameters. 
Moreover, we demonstrate GDD has better robustness against covariate shift.

\vspace{0.1in}
\noindent
\textbf{Notation:}
Vectors and matrices are written in bold lowercase and uppercase letters, respectively.
The $(i,j)$ element and the $j$-th column of a matrix $\mathbf{A}$ are denoted by $A_{i,j}$ and $\mathbf{a}_{j}$, respectively.
The $i$-th element in the vector $\mathbf{a}$ is denoted by $a_{i}$.
The square identity matrix of rank $N$ is denoted by $\mathbf{I}_N$, the $M$-by-$N$ zero matrix is denoted by $\mathbf{0}_{M,N}$, and the vector of all ones / zeros of length $N$ is denoted by $\mathbf{1}_N$ / $\mathbf{0}_N$, respectively.
Operator $\|\cdot\|_p$ denotes the $\ell$-$p$ norm.

\section{Preliminaries}
\label{sec:prelim}
\subsection{GSP Definitions}

A graph $\cG = (\cV, \cE, \W)$ is defined by a set $\cN = \{1, \ldots, N\}$ of $N$ nodes and an edge set $\cE = \{(i,j)\}$, where edge $(i,j) \in \cE$ has (positive or negative) weight $w_{i,j} = W_{i,j}$, for \textit{adjacency matrix} $\W \in \mathbb{R}^{N \times N}$. 
If edges are undirected, then $W_{i,j} = W_{j,i}$, and $\W$ is symmetric.
\textit{Degree matrix} $\D \in \mathbb{R}^{N \times N}$ is a diagonal matrix with diagonal entries $D_{i,i} = \sum_j W_{i,j}$. 
The combinatorial \textit{graph Laplacian matrix} $\L \in \mathbb{R}^{N \times N}$ \cite{ortega18ieee} is defined as
\begin{align}
\L \triangleq \D - \W .
\end{align}
$\L$ is provably \textit{positive semi-definite} (PSD) if all edge weights are non-negative, \ie, $W_{i,j} \geq 0, \forall i,j$ \cite{cheung18}. 
If $\cG$ contain self-loops, \ie, $\exists i, W_{i,i} \neq 0$, then the \textit{generalized graph Laplacian matrix} $\cL \in \mathbb{R}^{N \times N}$ is typically used:
\begin{align}
\cL \triangleq \D - \W + \text{diag}(\W) 
\end{align}
where $\text{diag}(\W)$ extracts the diagonal entries of $\W$. 

Real and symmetric Laplacian $\L$ (or $\cL$) can be eigen-decomposed to $\L = \V \bSigma \V^\top$, where $\V$ contains eigenvectors of $\L$ as columns, and $\bSigma = \text{diag}(\lambda_1, \ldots, \lambda_N)$ is a diagonal matrix with ordered eigenvalues $\lambda_1 \leq \lambda_2 \leq \ldots \leq \lambda_N$ along its diagonal.
The $k$-th eigen-pair $(\lambda_k,\v_k)$ is the $k$-th graph frequency and Fourier mode for $\cG$, respectively.
$\tilde{\x} = \V^\top \x$ is the \textit{graph Fourier transform} (GFT) of signal $\x$ \cite{ortega18ieee}, where $\tilde{x}_k = \v_k^\top \x$ is the $k$-th GFT coefficient for signal $\x$.

\subsection{Graph Laplacian Regularizer}

To regularize an ill-posed graph signal restoration problem like denoising \cite{pang17} or interpolation \cite{chen21}, GLR is popular due to its convenient quadratic form that is easy to optimize \cite{pang17}. 
Given a graph $\cG$ specified by (presumably PSD) graph Laplacian $\L$, GLR for a signal $\x \in \mathbb{R}^N$ is defined as
\begin{align}
\x^\top \L \x &= \sum_{(i,j) \in \cE} w_{i,j} (x_i - x_j)^2 = \sum_k \lambda_k \tilde{x}_k^2 .
\label{eq:GLR} 
\end{align}
Thus, minimizing GLR means promoting a signal with connected sample pairs $(x_i,x_j)$ that are similar.
In the spectral domain, it means promoting a \textit{low-pass} signal with energies $\tilde{x}_k^2$'s concentrated in low graph frequencies $\lambda_k$'s. 

If self-loops exist in graph $\cG$, instead of $\L$, one can define GLR alternatively using (presumably PSD) generalized Laplacian $\cL$ instead:
\begin{align}
\x^\top \cL \x &= \sum_{(i,j) \in \cE} w_{i,j} (x_i - x_j)^2 + \sum_i w_{i,i} x_i^2 .
\label{eq:GLR2}
\end{align}

\section{Algorithm Development}
\label{sec:formulate}
\subsection{Review of Theorem 1}

We first review Theorem 1 in \cite{viswarupan2023mixed}. 
Consider a linear denoiser $\bPsi \in \mathbb{R}^{N \times N}$ written as
\vspace{-0.05in}
\begin{align}
\x = \bPsi \y,
\label{eq:denoiser}
\end{align}
where $\x, \y \in \mathbb{R}^N$ are respectively the denoiser output and input. 
Next, consider a MAP optimization problem to denoise input $\y$, where GLR \eqref{eq:GLR2} \cite{pang17} is used as the signal prior:
\begin{align}
\min_\x \|\y - \x\|^2_2 + \mu \x^{\top} \cL \x .
\label{eq:MAP_denoise}
\end{align}
$\mu > 0$ is a weight parameter trading off the fildelity term and the prior.
Assuming that generalized graph Laplacian $\cL$ is PSD, \eqref{eq:MAP_denoise} is an unconstrained \textit{quadratic programming} (QP) problem with a convex objective, where the solution is
\begin{align}
\x^* = \left(\I_N + \mu \, \cL \right)^{-1} \y .
\label{eq:MAP_sol_denoise}
\end{align}
We call operator $(\I_N + \mu \cL)^{-1}$ a \textit{graph filter}, since objective in \eqref{eq:MAP_denoise} is regularized using a graph smoothness prior, GLR.
Note that coefficient matrix $\I_N + \mu \cL$ is provably \textit{positive definite} (PD) given $\cL$ is PSD, and thus invertible. 

We restate Theorem 1 \cite{viswarupan2023mixed} below.

\begin{theorem}
Denoiser $\bPsi$ \eqref{eq:denoiser} is the solution filter for the MAP problem \eqref{eq:MAP_denoise} if $\cL = \mu^{-1} (\bPsi^{-1} - \I_N)$, assuming matrix $\bPsi$ is non-expansive, symmetric, and PD. 
\label{thm:denoiser}
\end{theorem}
Theorem\;\ref{thm:denoiser} establishes, under mild conditions, a one-to-one mapping between any (pseudo-)linear denoiser $\bPsi$ and a corresponding graph filter \eqref{eq:MAP_sol_denoise} that is a solution to MAP problem \eqref{eq:MAP_denoise} specified by Laplacian $\cL$.
We leverage this theorem to initialize our proposed graph-based deep denoiser (GDD) next.

\subsection{Generalizing Bilateral Filter}
\label{subsec:generalize_BF}

We first select a trusted (pseudo-)linear\footnote{By pseudo-linear, we mean an operator $\bPsi(\x)$ that is a function of target signal $\x$, but once it is fixed at estimate $\bPsi(\x_o)$ in an optimization step, it is linear. BF \cite{tomasi98} is one example due to its edge weight dependency on signal $\x$.} denoiser $\bPsi$ with known denoising performance, such as BF \cite{tomasi98}.
BF\footnote{BF was interpreted as a specific graph filter in \cite{gadde13}. However, Theorem 1 in \cite{viswarupan2023mixed} is more general in that any (pseudo-)linear denoiser $\bPsi$ that is non-expansive, symmetric and PD can now be interpreted as a graph filter.} is a weighted averaging filter, where the filter weight $u_{i,j}$ between target pixel $i$ and a neighbor $j$ is
\begin{align}
u_{i,j} = \exp{\left(-\dfrac{\|\l_i - \l_j\|^2}{\sigma_l^2} \right)}
\exp{\left(-\dfrac{|x_i-x_j|^2}{\sigma_x^2} \right)} .
\label{eq:BF}
\end{align}
$\l_i \in \mathbb{R}^2$ and $x_i \in \mathbb{R}$ denote the 2D grid location and intensity for pixel $i$, respectively.
In \eqref{eq:BF}, the first term (called \textit{domain} filter in \cite{tomasi98}) reflects the pairwise Euclidean distance on the 2D grid, while the second term (called \textit{range} filter) reflects  the pairwise photometric distance. 
Note that $u_{i,j}$ is \textit{signal-dependent}, \ie, the filter weight $u_{i,j}$ used to restore signal $\x$ is itself a function of the signal $\x$. 

To generalize BF, beyond location $\l_i$ and intensity $x_i$, one can introduce other relevant features into the computation of filter weights $b_{i,j}$ \eqref{eq:BF}, similarly done in \cite{onuki16}.
Formally, we first define \textit{feature vector} $\f_i \in \mathbb{R}^K$, where the first three features are the horizontal/vertical coordinates of pixel $i$ on the 2D grid, $l^x_i$ and $l^y_i$, and pixel intensity $x_i$. 
The remaining features can be handcrafted features like horizontal/vertical image gradients, or learned features borrowed from the first layer of a pre-trained CNN.
We discuss our generalization in our experiments in Section\;\ref{subsec:exp_setup}.

Filter weight $b_{i,j}$ can now be more generally defined as
\begin{equation}
b_{i,j} = \exp \left( - (\f_i-\f_j)^\top \M (\mathbf{f_i-f_j}) \right)
\label{eq:BF_gen}
\end{equation}
where $\M \in \mathbb{R}^{K \times K} \succeq 0$ is a PSD \textit{metric matrix} \cite{yang22}. 
BF filter weights \eqref{eq:BF} is a special case of \eqref{eq:BF_gen}, where $\M = \text{diag}(1/\sigma_l^2, 1/\sigma_l^2,1/\sigma_x^2, 0, \ldots)$.
The importance of \eqref{eq:BF_gen} is that, given a set of pre-selected features to compose feature vectors $\{\f_i\}$, one can first initialize a general filter with BF weights for a specific metric $\M$, then learn from data a more general $\M$ for optimal performance. 
We discuss this in Section\;\ref{subsec:init_Laplacian}.

\subsubsection{Matrix Normalization}

In conventional BF, computed filter weights $\{b_{i,j}\}$ \eqref{eq:BF} in filter weight matrix $\B$ are normalized to $\{\bar{b}_{i,j}\}$, so that the sum of weights $\sum_j \bar{b}_{i,j}$ for filtering target pixel $i$---sum of entries in row $i$ of $\B$---adds up to $1$. 
Mathematically, this means pre-multiplying filter matrix $\B$ by diagonal matrix $\S^{-1}$, where $S_{i,i} = \sum_j b_{i,j}$. 
However, $\S^{-1} \B$ is not symmetric in general and hence does not satisfy the matrix symmetry condition for $\bPsi$ in Theorem\;\ref{thm:denoiser}.
One possibility is to perform the Sinkhorn-Knopp (SK) algorithm on $\B$ to make it doubly-stochastic \cite{sinkhorn67}. 
However, the iterative SK algorithm can be computationally expensive.

Instead, we perform a simpler symmetric normalization $\bPsi = \S^{-1/2} \B \S^{-1/2}$, which ensures that the normalized $\bPsi$ is symmetric and non-expansive, though it deviates slightly from standard BF normalization. 
We assume this procedure to normalize $\B$ to $\bPsi$ in the sequel.

\subsection{Initializing Graph Laplacian}
\label{subsec:init_Laplacian}

Given BF denoiser $\bPsi$, we leverage Theorem\;\ref{thm:denoiser} to initialize a graph Laplacian $\cL$ for an equivalent graph filter $(\I_N + \mu \cL)^{-1}$, from which we begin our data learning on $\cL$ for optimal performance.
However, computing $\cL = \mu^{-1}(\bPsi^{-1} - \I_N)$ according to Theorem\;\ref{thm:denoiser} requires a matrix inverse operation $\bPsi^{-1}$, which is $\cO(N^3)$ in the worst case.
Further, optimizing parameters of $\bPsi$ (entries in metric $\M$ in particular) end-to-end in the unrolled network via back-propagation (to be discussed in Section\;\ref{sec:unrolling}), when $\cL$ is a function of $\bPsi^{-1}$, is complicated. 

Instead, we approximate $\bPsi^{-1}$ using truncated \textit{Taylor Series Expansion} (TSE); this means that we can conveniently write $\cL$ as a finite-order polynomial of $\bPsi$.
Recall that the TSE of a differentiable function $f(x)$ at fixed point $x=s$ is
\begin{align}
f(x) = \sum_{k=0}^\infty \frac{f^{(k)}(s)}{k!} (x - s)^k .
\end{align}
Specializing the function to $f(x) = x^{-1}$, the TSE is
\begin{align}
x^{-1} = \sum_{k=0}^\infty \frac{a_k}{s^{k+1}} (x-s)^k ,   
\label{eq:TSE}
\end{align}
where the $k$-th TSE coefficient is $a_k = (-1)^k$. 

Assuming that matrix $\bPsi$ is non-expansive, symmetric, and PD, as required in Theorem\;\ref{thm:denoiser}, we can eigen-decompose $\bPsi$ and write $\bPsi = \U \mathbf{\Lambda} \U^\top$, where $\mathbf{\Lambda}$ and $\U$ are respectively the diagonal matrix containing $\bPsi$'s ordered eigenvalues along its diagonal, and the eigen-matrix with $\bPsi$'s eigenvectors as columns.
Using \eqref{eq:TSE} to express the function $f(\bPsi) = \bPsi^{-1}$, we write
\begin{align}
\bPsi^{-1} &= \U f(\mathbf{\Lambda}) \U^\top \\
&= \U \sum_{k=0}^\infty \frac{a_k}{s^{k+1}} \left( \mathbf{\Lambda} - s\I_N \right)^k \U^\top  \\ 
&= \sum_{k=0}^\infty \frac{a_k}{s^{k+1}} \left( \bPsi - s\I_N \right)^k .  
\label{eq:TSR}
\end{align}

Hence, from Theorem\;\ref{thm:denoiser} we can express the Laplacian matrix $\cL$ as a polynomial of $\bPsi$, using the truncated TSE in \eqref{eq:TSR} at degree $K$:
\begin{align}
\cL \approx \mu^{-1} \left(
 \sum_{k=0}^K \frac{a_k}{s^{k+1}} \left( \bPsi - s\I_N \right)^k  - \I_N \right) .
 \label{eq:Lknot}
\end{align}
Thus, initializing $\cL$ can be done by using a finite-order polynomial of $\bPsi$.

\section{Algorithm Unrolling}
\label{sec:unrolling}
\subsection{Laplacian Sub-network}

\begin{figure}[t]
\begin{center}
\includegraphics[width=0.9\linewidth]{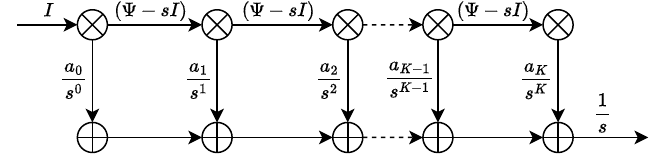}
\end{center}
\vspace{-0.1in}
\caption{Feed-forward sub-network to implement a truncated TSE of $f(\bPsi) = \bPsi^{-1}$ according to \eqref{eq:TSR}. TSE coefficients $a_k$ for each $\bPsi_i$ 
are learned through end-to-end training. }
\label{fig:Laplacian_subnet}
\end{figure}

We implement the computation of Laplacian $\cL$ in \eqref{eq:Lknot} as a \textit{feed-forward network} (FFN).
In particular, the truncated TSE approximation of $\bPsi^{-1}$ is implemented in a network shown in Fig.\;\ref{fig:Laplacian_subnet}. 
We learn TSE coefficients $a_k$ for each $\bPsi_k$ in the sequence of length $K$.
Note that the coefficients $a_k$ are initialized to be the original derived TSE coefficient values, \ie, $a_k = (-1)^k$, thus guaranteeing a minimum baseline performance before we tune parameters for performance gain.

\subsection{Unrolling Conjugate Gradient Descent}

Having $\cL$ expressed as a finite-order polynomial of $\bPsi$, instead of computing graph filter $(\I_N + \mu \cL)^{-1}$ via a matrix inverse in \eqref{eq:MAP_sol_denoise}, we solve for solution $\x^*$ of the linear system
\begin{align}
(\I_N + \mu \cL) \x^* = \y
\label{eq:cgdeq}
\end{align}
by unrolling \textit{conjugate gradient} (CG) \cite{stiefel52, saad2003iterative} also into a FFN, and thus amenable to end-to-end learning. 
The iterative CG algorithm have the following update process, corresponding to four intermediate states: $\v_{k}$, $\x_{k}$, $\r_{k}$ and $\p_{k}$,
\begin{align}
    \v_{k+1} &= (\I_N + \mu \cL) \p_{k} \\
    \x_{k+1} &= \x_{k} + \alpha_k \p_{k} \\
    \r_{k+1} &= \y - (\I_N + \mu \cL) \x_{k+1} = \r_{k} -  \alpha_k \v_{k+1} \\
    \p_{k+1} &= \r_{k+1} + \beta_k \p_{k}
\end{align}
where $\alpha_k$ and $\beta_k$ can be understood as \textit{learning rate} and \textit{gradient momentum}. 
\begin{align}
    \alpha_k &= \frac{\r_{k}^\top\r_{k}}{\p_{k}^\top\v_{k+1}} 
    \label{eq:cgparams1}
    \\
    \beta_k &= \frac{\r_{k+1}^\top\r_{k+1}}{\r_{k}^\top\r_{k}} .
\label{eq:cgparams2}
\end{align}

\subsection{Loss Function}

Given noisy and ground truth image patches $\{\y_k,\x_k\}$, we learn parameters $\theta$ of differentiable function using the following square error loss function:
\begin{align}
\min_\theta l(\{y_k,\x_k\}) = \min_\theta \sum_{k=1}^K \| \x_k - f_\theta(\y_k) \|^2_2  .
\end{align}
Specifically, the set of parameters we learn includes $\alpha_k$ \eqref{eq:cgparams1} and $\beta_k$ \eqref{eq:cgparams2} in the CG algorithm, TSE coefficients $\{a_k\}_{k=0}^K$ in \eqref{eq:Lknot}, and metric matrix $\M$ entries in \eqref{eq:BF_gen}.

\section{Experiments}
\label{sec:results}
\subsection{Experimental Setup}
\label{subsec:exp_setup}

To test our proposed GDD unrolled network against competing schemes, we used TAMPERE17 dataset \cite{dset} consisted of 300 color images of resolution $512 \times 512$. 
We contaminated dataset with additive white Gaussian noise (AWGN) of chosen noise standard deivation (SD) and randomly split the dataset into test and training sets with 100 and 200 images, respectively. 
Then, we partitioned each image into non-overlapping $N\times N$ patches. 
During training, we inputted noisy patches in the training set through the network, and computed MSE between network output and corresponding ground truth patch. 
We learned CG parameters, TSE coefficients, and metric matrix $\M$ entries.
For initialization, we approximated $\bPsi^{-1}$ using the sub-network shown in Fig.\;\ref{fig:Laplacian_subnet}, and then learned parameters deviating from initialized values.

\begin{figure}[t]
\begin{center}
\includegraphics[width=0.9\linewidth]{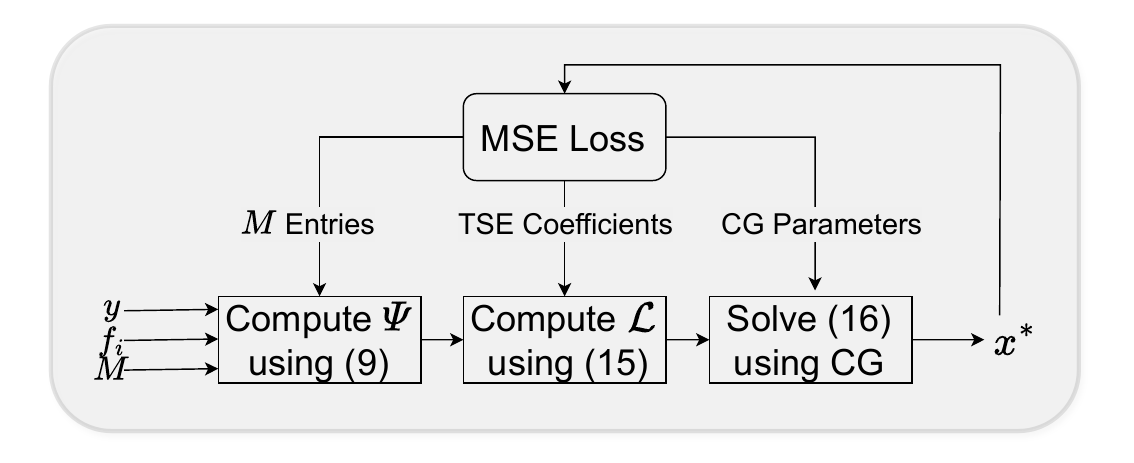}
\end{center}
\vspace{-0.2in}
\caption{Overview of proposed network architecture, here $y$ is the noisy patch, $f_i$'s are hand-crafted features and $M$ is initialized metric matrix.}
\label{fig:overal_structure}
\end{figure}
For feature vectors $\{\f_i\}$ in \eqref{eq:BF_gen}, we considered five pre-computed features. 
The first three features were pixel horizontal/vertical coordinates and input pixel intensity---the same as BF.
The last two features were horizontal/vertical image gradients. 
Given that we learned a relatively small matrix $\M$ per layer, $K$ TSE coefficients ($K=10$), and two parameters for each CG iteration (we set CG iterations to be $15$), our denoiser required training of very few parameters compared to conventional DL-based denoisers. We empirically chose $s=1$ in \eqref{eq:Lknot}. Weight parameters $\mu^{-1}$ in \eqref{eq:Lknot} and $\mu$ in \eqref{eq:cgdeq} were chosen to be the same.
An overview of network architecture is shown in Fig.\;\ref{fig:overal_structure}. 
At the initialization stage, noisy patch $\y$, hand-crafted features $\f_i$, and initialized metric matrix $\M$ are inputted to the network. 
BF denoiser ($\bPsi$) then can be calculated using \eqref{eq:BF_gen}. 
Next, we approximated $\cL$ using \eqref{eq:Lknot} through FFN shown in Fig.\;\ref{fig:Laplacian_subnet}. 
Having an approximation of $\cL$ in hand, we solved \eqref{eq:MAP_sol_denoise} via \eqref{eq:cgdeq} using the unrolled CG algorithm. 
Parameters of each layer were learned in an end-to-end manner through back-propagation.
We trained our network for 20 epochs using $64\times64$ patches and batch size 3 using Adam optimizer with learning rate 0.001.

\subsection{Experimental Results}

\begin{figure}[H]
\includegraphics[width=\linewidth]{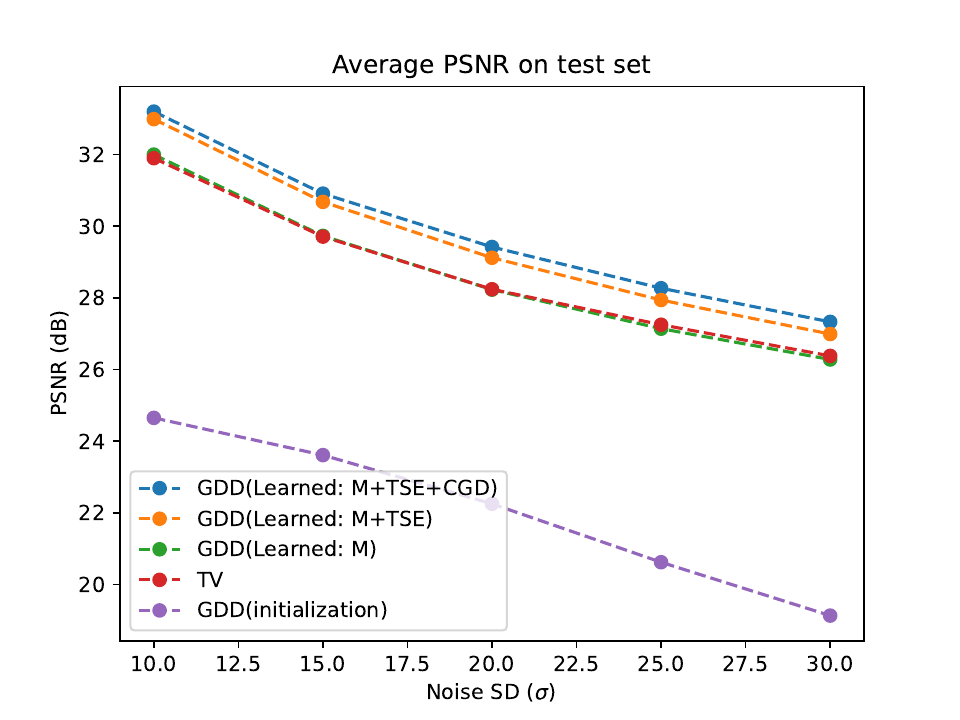}
\vspace{-0.3in}
\caption{PSNR versus noise SD for variants of trained unrolled network, initialized network, and total variation.} 
\label{fig:plot}
\end{figure}

Fig.\;\ref{fig:plot} shows the average resulting denoised image quality in PSNR for conventional BF (GDD initialization), Total Variation (TV), and variants of our proposed GDD unrolled network at different noise variances for a set of test images.
Note that we restricted our comparisons to competitors that are also \textit{local}, \ie, no non-local similar patch search, which generally necessitates a larger processing memory requirement and higher computation complexity, resulting in non-real-time processing.
We observe that our proposed GDD outperformed BF at all noise levels, at one point by more than 8.53dB. 
Our proposed GDD also outperformed TV at all noise levels , at one point by more than $1.28$dB.

We observe also that by training more parameters in GDD, performance generally improved.
Specifically, training entries in the metric matrix $\M$ improved performance over initialization noticeably, then training in addition TSE coefficients brought further gain, and training CG parameters improved performance even more.
The amount of improvement from employing three features to more features was less dramatic, but as discussed in Section\;\ref{subsec:generalize_BF}, an implementation generalizing from BF features to pre-trained features from a CNN first layer can potentially induce more gain. 

\begin{figure}
\centering
      \subfloat[\centering Noisy image (28.29dB)]{\includegraphics[width=.2\textwidth]{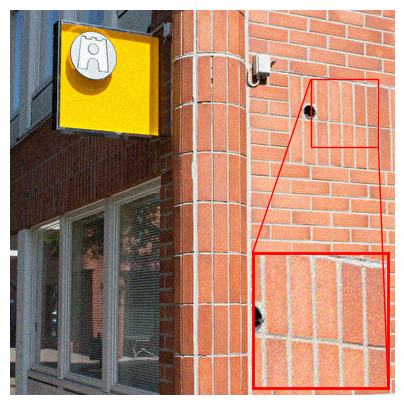}}
      \hspace{0.1cm}
      \subfloat[\centering Bilateral Filter (29.91dB)]{\includegraphics[width=.2\textwidth]{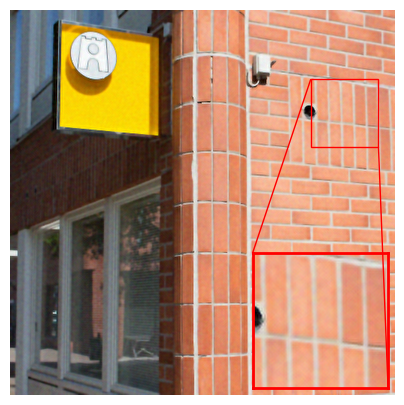}}
      \hspace{0.1cm}
      \subfloat[\centering TV (33.13dB)]{\includegraphics[width=.2\textwidth]{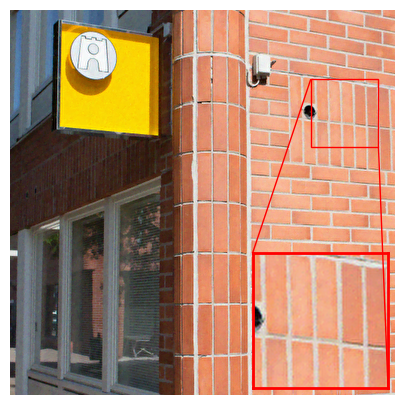}}
      \hspace{0.1cm}
      \subfloat[\centering Proposed (34.78dB)]{\includegraphics[width=.2\textwidth]{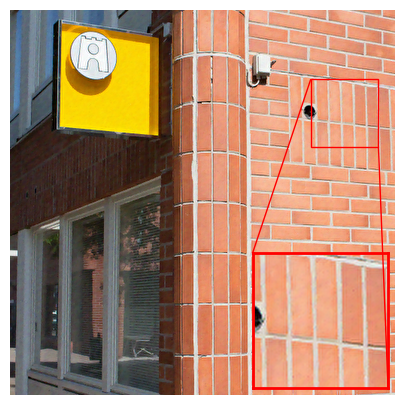}}
  \vspace{-0.1in}
  \caption{Examples of noisy and denoised \texttt{wall} images by competing methods and their quality in PSNR.}
  \label{fig:denoised_images}
\end{figure}

\begin{figure}
\centering
      \subfloat[\centering Noisy image (28.25dB)]{\includegraphics[width=.2\textwidth]{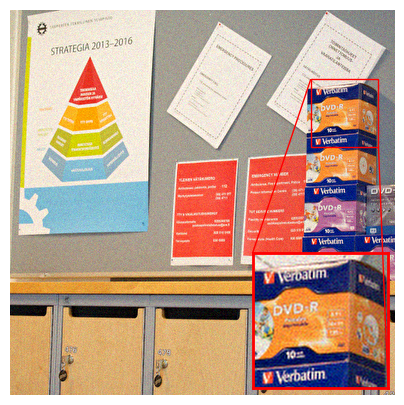}}
      \hspace{0.1cm}
      \subfloat[\centering Bilateral Filter (30.46dB)]{\includegraphics[width=.2\textwidth]{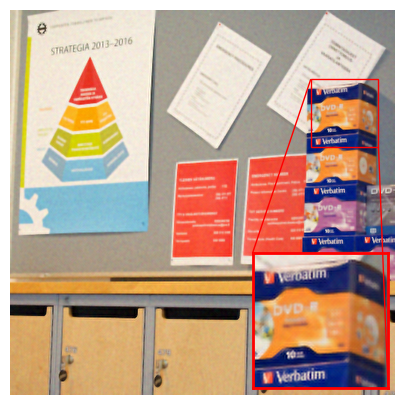}}
      \hspace{0.1cm}
      \subfloat[\centering TV (33.49dB)]{\includegraphics[width=.2\textwidth]{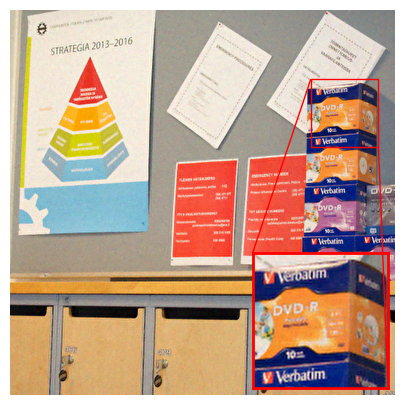}}
      \hspace{0.1cm}
      \subfloat[\centering Proposed (35.65dB)]{\includegraphics[width=.2\textwidth]{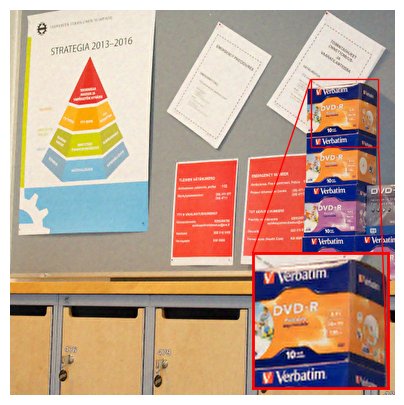}}
  \vspace{-0.1in}
  \caption{Examples of noisy and denoised \texttt{office} images by competing methods and their quality in PSNR.}
  \label{fig:denoised_images2}
\end{figure}

Fig.\;\ref{fig:denoised_images} and \ref{fig:denoised_images2} show examples of noise-corrupted images and denoised images by BF, TV and our trained GDD.
In these cases, we observe that our trained GDD outperformed BF by up to $4.84$dB in PSNR, and outperformed TV by up to $2.03$dB.
Visually, we see that GDD produced denoised images while maintaining sharpness compared to BF and TV.

\begin{figure}[!ht]
\centering
      \subfloat[\centering GDD (28.29dB)]{\includegraphics[width=.2\textwidth]{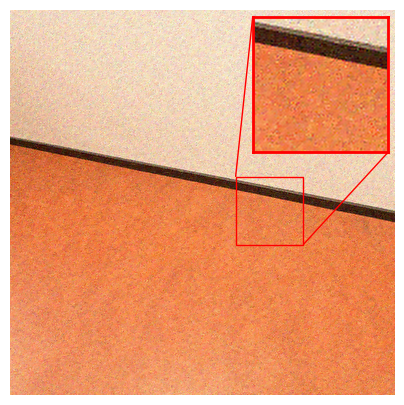}}
      \hspace{0.1cm}
      \subfloat[\centering DnCNN (27.04dB)]{\includegraphics[width=.2\textwidth]{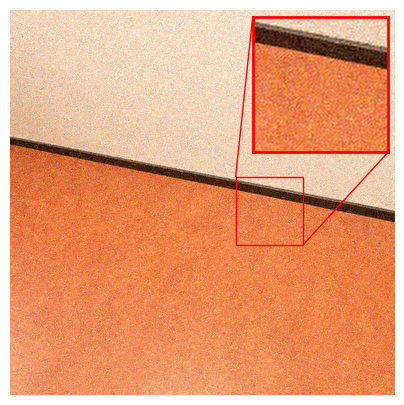}}
      \hspace{0.1cm}
  \vspace{-0.1in}
  \caption{Results in statistical mismatch and small dataset training. Training on $\sigma=10$ and testing on $\sigma=20$.}
  \label{fig:denoised_images3}
\end{figure}

To demonstrate our proposed model's robustness against \textit{covariate shift}---statistical mismatch between test and training set---we first trained our model using 200 randomly selected images from the dataset with artificial noise $\sigma=10$ and tested our model using 100 other random images with a different noise SD.
Then, we trained DnCNN \cite{zhang17} with exactly the same small training set and tested it under the same conditions. 
The resulting denoised image PSNR versus testing data noise SD is shown in Table\;\ref{fig:covartable}. 
We observe that GDD was competitive with DnCNN for all noise SDs given the small training dataset. 
Specifically, GDD outperformed DnCNN at all noise SDs greater than $20$. 
See Fig.\;\ref{fig:denoised_images3} for examples of denoised images at $\sigma = 20 $ by the two schemes. GDD outperformed DnCNN by around 1.25dB in this case of small training dataset and covariate shift. 
This demonstrates the effectiveness and robustness of our unrolled network that requires training of very few parameters from a small training dataset, compared to DnCNN.

\begin{figure}

    \centering
    \begin{footnotesize}
\begin{tabular}{c|ccccc}
\hline
\multirow{2}{*}{Method} & \multicolumn{5}{c}{PSNR on Test set for specified $\sigma$}                        \\
                        & $\sigma=10$    & $\sigma=15$    & $\sigma=20$    & $\sigma=25$    & $\sigma=30$    \\ \hline
DnCNN                   & \textbf{35.24} & \textbf{30.45} & 25.82          & 22.90          & 20.87          \\
GDD                     & 33.23          & 29.79          & \textbf{26.19} & \textbf{23.22} & \textbf{20.89} \\ \hline
\end{tabular}
\end{footnotesize}
\captionof{table}{PSNR versus noise SD for test set. Both models were trained on small dataset with noise SD $\sigma=10$.}
    \label{table:t2}
\label{fig:covartable}
\end{figure}

\section{Conclusion}
\label{sec:conclude}
We propose a general framework to construct an interpretable graph-based deep denoiser (GDD) involving three major steps.
First, we choose a trusted (pseudo-)linear denoiser $\bPsi$ with known good denoising performance, such as the bilateral filter.
Second, we approximate the corresponding graph Laplacian matrix $\cL = \mu^{-1} (\bPsi^{-1} - \I_N)$, leveraging Theorem\;1 in \cite{viswarupan2023mixed}, stating a one-to-one mapping between denoiser $\bPsi$ and an equivalent graph filter for a MAP denoising problem with a graph Laplacian regularizer (GLR) as prior, via truncated Taylor series expansion (TSE).
Finally, given $\cL$ we solve a linear system to compute the filter solution by unrolling a conjugate gradient (CG) algorithm. 
The resulting unrolled network is fully interpretable, and easily initialized using trusted $\bPsi$ and thus ensuring a baseline performance.
It also requires tuning of very few parameters using a small dataset.
Experimental results show that our data-optimized unrolled network outperformed the untrained initialized network and is competitive with competing methods, and is more robust to covariate shift.
Deep network training using a small dataset for few parameters is important for future algorithm design, given the worsening global climate change and environmentally costly energy consumption.  

\begin{small}
\bibliographystyle{IEEEbib}
\bibliography{ref2}
\end{small}

\end{document}